\documentclass[a4paper,fleqn]{cas-dc}

\usepackage[numbers]{natbib}

\def\tsc#1{\csdef{#1}{\textsc{\lowercase{#1}}\xspace}}
\tsc{WGM}
\tsc{QE}
\tsc{EP}
\tsc{PMS}
\tsc{BEC}
\tsc{DE}

\usepackage{graphicx}
\usepackage{multirow}

\usepackage{amsmath}

\usepackage{diagbox} 
\usepackage{tabularx}
\usepackage{times}
\usepackage{epsfig}
\usepackage{amssymb}
\usepackage{verbatim}
\usepackage{mathrsfs}

\usepackage{algorithm}
\usepackage{algorithmic}
\usepackage{color}

\usepackage{amsthm}

\usepackage{tikz}
\newcommand*{\circled}[1]{\lower.7ex\hbox{\tikz\draw (0pt, 0pt)%
		circle (.4em) node {\makebox[0.6em][c]{\small #1}};}}

\usepackage{subfig}

\begin{document}
\let\WriteBookmarks\relax
\def\floatpagepagefraction{1}
\def\textpagefraction{.001}
\shorttitle{A GAN-Based Input-Size Flexibility Model for Single Image Dehazing}
\shortauthors{Shichao Kan et~al.}

\title [mode = title]{A GAN-Based Input-Size Flexibility Model for Single Image Dehazing} 


\tnotetext[1]{This work was supported in part by the National Key R\&D Program of China under Grant 2021YFE0110500; in part by the National Natural Science Foundation of China under Grant 61872034, Grant 62011530042, and Grant 62062021; in part by the Beijing Municipal Natural Science Foundation under Grant 4202055; inpart by the Fundamental Research Funds for the Central Universities 2021QY002.}

\author[1,2,3]{Shichao Kan}[style=chinese,
                         orcid=0000-0003-0097-6196]
\fnmark[1]
\ead{kanshichao@csu.edu.cn}

\address[1]{School of Computer Science and Engineering, Central South University, 410083, Changsha, Hunan, China}
\address[2]{Institute of Information Science, Beijing Jiaotong University, Beijing 100044, China}
\address[3]{Beijing Key Laboratory of Advanced Information Science and Network Technology, Beijing 100044, China}

\author[2,3]{Yue Zhang}[style=chinese]
\fnmark[1]
\ead{17112065@bjtu.edu.cn}
\author[2,3]{Fanghui Zhang}[style=chinese]
\ead{18112013@bjtu.edu.cn}
\author[2,3]{Yigang Cen}[style=chinese]
\cormark[1]
\ead{ygcen@bjtu.edu.cn}







\cortext[cor1]{Corresponding author}
\fntext[fn1]{The first two authors (Shichao Kan and Yue Zhang) contribute equally.}


\begin{abstract}
Image-to-image translation based on generative adversarial network (GAN) has achieved state-of-the-art performance in various image restoration applications. Single image dehazing is a typical example, which aims to obtain the haze-free image of a haze one. This paper concentrates on the challenging task of single image dehazing. Based on the atmospheric scattering model, a novel model is designed to directly generate the haze-free image. The main challenge of image dehazing is that the atmospheric scattering model has two parameters, i.e., transmission map and atmospheric light. When they are estimated respectively, the errors will be accumulated to compromise the dehazing quality. Considering this reason and various image sizes, a novel input-size flexibility conditional generative adversarial network (cGAN) is proposed for single image dehazing, which is input-size flexibility at both training and test stages for image-to-image translation with cGAN framework. A simple and effective U-connection residual network (UR-Net) is proposed to combine the generator and adopt the spatial pyramid pooling (SPP) to design the discriminator. Moreover, the model is trained with multi-loss function, in which the consistency loss is a novel designed loss in this paper. Finally, a multi-scale cGAN fusion model is built to realize state-of-the-art single image dehazing performance. The proposed models receive a haze image as input and directly output a haze-free one. Experimental results demonstrate the effectiveness and efficiency of the proposed models.

\end{abstract}



\begin{keywords}
generative adversarial network \sep image dehazing \sep image restoration
\end{keywords}

\maketitle

\section{Introduction}

Haze removal \cite{1He0T11} is a classical ill-posed image restoration problem, which plays an important role in intelligent transportation systems, e.g., object detection under haze conditions \cite{2LiPWXF18,4LiRFTFZW19,3abs-1807-00202}. Haze is defined as some particles such as dust that obscure the clarity of the atmosphere. Dehazing is to remove the veil of haze from a haze image and restore a corresponding haze-free image. In recent years, because the development of deep learning has greatly improved the performance of image processing compared with non-learning-based technology, the problem of dehazing attracts more and more attentions in image restoration research community. Various image dehazing methods based on deep learning technology have been proposed, including: (1) Generating medium transmission map \cite{5CaiXJQT16} or haze-free image \cite{6LiPWXF17,3abs-1807-00202,6abs-1906-04334} by a convolutional neural network (CNN); (2) generating transmission map \cite{7RenLZPC016} or haze-free image \cite{9LiuXAWC19,9abs-1805-03305,80004SP18} based on encoder-decoder structure without adversary training; (3) reconstructing haze-free image with paired image-to-image translation models based on generative adversary network (GAN) \cite{12LiPLT18,14Qu,13Ren0ZPC0018,100004P18a,11ZhuPCLL18}; (4) reconstructing haze-free image with unpaired image-to-image translation models based on cycle GAN (CGAN) \cite{16EnginGE18,17abs-1902-01374,15YangXL18}.

In order to directly generate medium transmission map, \cite{5CaiXJQT16} and \cite{7RenLZPC016} proposed an end-to-end learnable CNN model. To generate haze-free image from a haze one via an end-to-end manner,  \cite{6LiPWXF17},\cite{3abs-1807-00202} and \cite{6abs-1906-04334} proposed light-weight and fast CNNs. \cite{9LiuXAWC19}, \cite{9abs-1805-03305} and \cite{80004SP18}  incorporated some modern technologies into CNNs based on encoder-decoder structure. Usually, the real of image restoration is sub-optimal based on these models. In order to merge GAN \cite{29GoodfellowPMXWOCB14} and image dehazing, supervised learning model with paired and unpaired samples based on adversary training are developed. \cite{12LiPLT18}, \cite{14Qu}, \cite{13Ren0ZPC0018}, \cite{100004P18a} and \cite{11ZhuPCLL18} are GAN-based end-to-end learnable models that trained with paired synthetic dataset, while \cite{15YangXL18}, \cite{16EnginGE18} and \cite{17abs-1902-01374} are cycle-consistency models that trained with unpaired training dataset.

From these deep learning-based methods, we can see that: (1) Because the end-to-end dehazing models \cite{6LiPWXF17,12LiPLT18,9LiuXAWC19,13Ren0ZPC0018,9abs-1805-03305,100004P18a,80004SP18,6abs-1906-04334,11ZhuPCLL18} can directly generate the haze-free image without additional parameter estimation, they are generally more efficient than non-end-to-end dehazing models \cite{5CaiXJQT16,7RenLZPC016}; (2) Due to down-sampling and up-sampling process are not used before and after image dehazing with input-size flexibility models \cite{6LiPWXF17,9LiuXAWC19}, the information loss can be minimized throughout the restoration process. Thus, images generated with input-size flexibility models have a better visual effect than images generated with input-size fixed models \cite{13Ren0ZPC0018,100004P18a,80004SP18,6abs-1906-04334}; (3) Because the paired samples have definitive supervised information, the training of the network can be truly supervised when the paired samples are used. Thus, paired image-to-image translation models \cite{12LiPLT18,14Qu,13Ren0ZPC0018,100004P18a,11ZhuPCLL18} are usually more effective than unpaired image-to-image translation models \cite{16EnginGE18,17abs-1902-01374,15YangXL18}; (4) Various works \cite{12LiPLT18,14Qu,13Ren0ZPC0018,11ZhuPCLL18} focus on exploring single image dehazing with GAN-based models and achieve promising performance. Considering these properties, we propose an end-to-end input-size flexibility conditional generative adversarial network (cGAN) for single image dehazing. The proposed model can not only remove the haze as much as possible but also preserve the clear content of an image.

The method proposed in this paper has obtained the state-of-the-art results on the datasets of the intelligent traffic video image enhancement processing competition\footnote{http://icig2019.csig.org.cn/?page\_id=328} of ICIG 2019 and the more large scale REalistic Single Image DEhazing (RESIDE) dataset \cite{4LiRFTFZW19} for single image dehazing. The performance improvement primarily comes from the input-size flexibility training and test, multi-loss supervised training and the designed end-to-end framework.

Our works have the following contributions. 
\begin{itemize}
\item
	An end-to-end input-size flexibility cGAN model is proposed for single image dehazing. The size of feature map in each layer of the generator is automated calculated based on the size of the input image. Based on our model, input-size flexibility mode can be applied to both adversary training and test stages and the image dehazing performance can be improved greatly. 
\end{itemize}
\begin{itemize}
	\item
	In our framework, a UR-Net structure is designed based on the popular U-Net \cite{18Tang} structure and residual learning \cite{19HeZRS16}, which is simple and effective. The generator is the iteration of UR-Net between two adjacent convolutional layers. Moreover, in order to realize input-size flexibility adversary training, the spatial pyramid pooling (SPP) \cite{21HeZR015} structure is embedded into the discriminator. 
\end{itemize}
\begin{itemize}
	\item 
    Training with multi-loss functions is also an important part of our framework. We proposed a novel consistency loss to keep the transformation consistency between the generated dehazing image and the real input image, and combined adversary loss, $L_1$ loss, the structural similarity (SSIM) loss and a new peak signal to noise ratio (PSNR) loss to train our network. The effectiveness of these loss functions is verified by ablation studies.
\end{itemize}

The rest of this paper is organized as follows. In Section \ref{se:2}, related works about learning-based single image dehazing are reviewed. The idea, framework and details of the proposed input-size flexibility cGAN for single image dehazing are presented in Section \ref{se:3}. In Section \ref{se:4}, datasets,  evaluation metrics and the experimental results are  presented. Section \ref{se:5} concludes the paper.

\section{Related Work}
\label{se:2}

Single image dehazing is a difficult vision task and has a long research history. Traditional single image dehazing methods are based on the handcrafted priors \cite{40GaoHLGP19}, e.g., dark channel prior \cite{1He0T11,sivp-Salazar-Colores18,ejivp-Salazar-Colores19}, color attenuation prior \cite{22ZhuMS15} and non-local prior \cite{23BermanTA16,24LiuGHL18}, which are usually simple and effective for many scenes. However, prior-based methods are limited when describing specific statistics. In recent few years, learning-based methods are becoming popular because they can overcome the limitations of specific priors \cite{SPIC-DingLWF21,SPIC-NairS21}. We also oriented to study learning-based single image dehazing in this paper. Here, works related to them are reviewed in detail, including learning-based dehazing without and with GAN methods, respectively.

\subsection{Learning-based Dehazing Without GAN}

Learning-based dehazing methods become more and more popular since the learning idea was proposed by Tang {\it et al.}\cite{25TangYW14}. The original idea was learning a regression model based on random forests from prior-based haze-relevant features, such as dark channel \cite{1He0T11}, local max contrast \cite{26Tan08}, hue disparity \cite{27AncutiAHB10}, and local max saturation \cite{25TangYW14}. Subsequently, more powerful learning dehazing models were proposed, especially CNN-based end-to-end learning methods. Song {\it et al.} \cite{28SongLWC18} proposed a ranking CNN to capture the statistical and structural attributes of hazy images, simultaneously. However, it is not an end-to-end learning system. Cai {\it et al.} \cite{5CaiXJQT16} proposed an end-to-end learning system to directly generate a medium transmission map, which is based on the CNN framework and called DehazeNet. Ren {\it et al.} \cite{7RenLZPC016} proposed a coarse-to-fine multi-scale CNN (MSCNN) model to predict transmission maps. Although the two models can be learned via an end-to-end manner, they are not end-to-end dehazing models.

In 2017, Li {\it et al.} \cite{6LiPWXF17} proposed a light-weight, effective and fast end-to-end learning model for image dehazing, called AOD-Net, which can directly generate a haze-free image from a haze one. Since then, the end-to-end dehazing idea is favored by researchers. Based on the AOD-Net framework, Liu {\it et al.} \cite{3abs-1807-00202} investigated various loss functions and demonstrated that training with perception-driving loss can further boost the performance of dehazing. Zhang {\it et al.} \cite{80004SP18} proposed a multi-scale image dehazing method using a perceptual pyramid deep network based on an encoder-decoder structure with a pyramid pooling module. In this model, the designed network is based on dense blocks \cite{8HuangLMW17} and residual blocks \cite{19HeZRS16}, the perceptual loss is also incorporated into the training process. Xu {\it et al.} \cite{9abs-1805-03305} proposed an instance normalization unit and embedded it into the VGG-based \cite{9SimonyanZ14a} U-Net \cite{18Tang} with an encoder-decoder structure. Liu {\it et al.} \cite{9LiuXAWC19} proposed a generic model-agnostic CNN (GMAN) for signal image dehazing, which is based on the fully convolutional idea and is not rely on the atmosphere scattering model. Both Xu {\it et al.} and Zhang {\it et al.} are based on the mean squared error (MSE) and VGG-feature-based perceptual loss to train the network. Recently, Zhang and Tao \cite{6abs-1906-04334} proposed a fast and accurate multi-scale end-to-end dehazing network called FAMED-Net, which is lightweight and computationally efficient.

Inspired by the success of these models, our proposed framework is based on the U-Net structure and residual learning, which is also an end-to-end dehazing one. Different from the previous idea, our network is designed for generalized image restoration, especially for different sizes of images, which can accept input images of any size during both training and test processes.

\subsection{Learning-based Dehazing With GAN}

The idea of GAN was first proposed in \cite{29GoodfellowPMXWOCB14}, which is designed to synthesize realistic images via an adversarial process. Latter, it is widely extended to a variety of image generation tasks, such as conditional image generation \cite{30MirzaO14}, paired image-to-image translation \cite{31IsolaZZE17}, unpaired image-to-image translation \cite{32ZhuPIE17}, etc. Now, it is also becoming popular in single image dehazing. Zhang and Patel \cite{100004P18a} proposed to jointly learn the transmission map, atmospheric light, and dehazing based on GAN framework, which is called densely connected pyramid dehazing network (DCPDN) and is an end-to-end single image dehazing model. Zhu {\it et al.} \cite{11ZhuPCLL18} formulated the atmospheric scattering model into a GAN framework and proposed a DehazeGAN, which can be used to learn the global atmospheric light and the transmission coefficient simultaneously. In order to generate realistic clear images, Li {\it et al.} \cite{12LiPLT18} directly estimates the haze-free image based on an end-to-end trainable cGAN with encoder-decoder architecture. Ren {\it et al.} \cite{13Ren0ZPC0018} adopted a fusion-based strategy to fuse three inputs from an original hazy image and proposed an end-to-end gated fusion network (GFN) for single image dehazing, which is trained with MSE and adversarial loss. Qu {\it et al.} \cite{14Qu} directly generate a haze-free image from a haze one without the physical scattering model, which is called enhanced pix2pix dehazing network (EPDN), and multi-loss function optimization idea is also used to train the network, including adversarial loss. All of these models are based on paired image-to-image translation framework.

Moreover, the unpaired image-to-image translation framework can be also found in single image dehazing. Yang {\it et al.} \cite{15YangXL18} proposed an end-to-end disentangled dehazing network to generate a haze-free image based on unpaired supervision. Engin {\it et al.} \cite{16EnginGE18} completed the dehazing task based on unpaired supervision, which did not rely on the atmospheric scattering model and trained by combining cycle-consistency and perceptual losses. Liu {\it et al.} \cite{17abs-1902-01374} developed an end-to-end learning system that uses unpaired fog and fog-free training images to generate a fog-free image, which also uses adversarial discriminators and cycle-consistency losses to train the whole framework. The advantage of unpaired supervision training is that the training process does not need to rely on synthetic dataset, because unpaired samples are easy to obtain. However, because these frameworks do not rely on the paired training data, the performance to restore realistic images is limited.

Therefore, our designed framework is based on paired cGAN, which is also incorporating multi-loss function optimization in it.

\section{Input-Size Flexibility Conditional Generative Adversarial Network}
\label{se:3}

Most of the previous single image dehazing models are based on the atmospheric scattering model, which tends to estimate the parameters of transmission map and atmospheric light. However, parameter estimation usually introduces estimation errors, which reduces the quality of restoration image. We thus develop an end-to-end and image-to-image translation model for single image dehazing, which is independent of the atmospheric scattering model and there is no additional parameter estimation. The proposed model directly produces a haze-free image from a haze one and is input-size flexibility at both training and test stages. In the following, the atmospheric scattering model is first analyzed and then each component of our proposed input-size flexibility conditional generative adversarial network is presented, respectively, i.e., generator, discriminator and loss functions.

\subsection{The Analysis of Atmospheric Scattering Model}
\label{se:3-1}

The famous atmospheric scattering model \cite{33NarasimhanN03} can be formulated as follows:
\begin{equation}
I_{re}(x) = J_{re}(x)t(x)+\alpha(1-t(x)),
\label{eq:1}
\end{equation}
where $I_{re}(x)$ is the real haze image that need to be restored, $J_{re}(x)$ is the expected haze-free image that could be recovered from $I_{re}(x)$, $t(x)$ is the medium transmission map, $\alpha$ is the global atmospheric light and $x$ is the indexes of the pixels corresponding to an image ($I_{re}$, $J_{re}$ and $t$). In real tasks, only $I_{re}(x)$ of Eq.(\ref{eq:1}) is known, the other three variables are unknown. Because the final goal is to estimate $J_{re}(x)$, thus if $t(x)$ and $\alpha$ can be estimated, then one can directly obtain the $J_{re}(x)$ according the following formula:
\begin{equation}
J_{re}(x) = \frac{1}{t(x)}I_{re}(x)+\alpha(1-\frac{1}{t(x)})
\label{eq:2}
\end{equation}

However, estimating $t(x)$ is a complex task because $t(x)$ is related with both the distance $d(x)$ from the scene point to the camera and the scattering coefficient $\beta$ of the atmosphere, which can be formulated as follows:
\begin{equation}
t(x)=e^{-\beta d(x)}
\label{eq:3}
\end{equation}

Moreover, there will always exist an error in the estimation of each parameter. Suppose that $\delta_1$ and $\delta_2$ are the average estimation errors of parameters $t$ and $\alpha$, respectively. When Eq.(\ref{eq:2}) is used to obtain a haze-free image, if the total average estimated error is $\delta$, then we have:
\begin{equation}
\delta=\delta_1+\delta_2+\delta_1\cdot\delta_2
\label{eq:4}
\end{equation}

From Eq.(\ref{eq:4}), only both $\delta_1 \rightarrow 0$ and $\delta_2 \rightarrow 0$, we can obtain $\delta \rightarrow 0$. When the estimating parameters are more than one in a system, the estimated error of each parameter is usually difficult to control simultaneously. In order to estimate them by an end-to-end manner, a new framework with a novel consistency loss (Eq.(\ref{eq:consistency})) is designed. Next, we will analyze it.

\begin{figure*}
	\centering
	\includegraphics[width=0.8\linewidth]{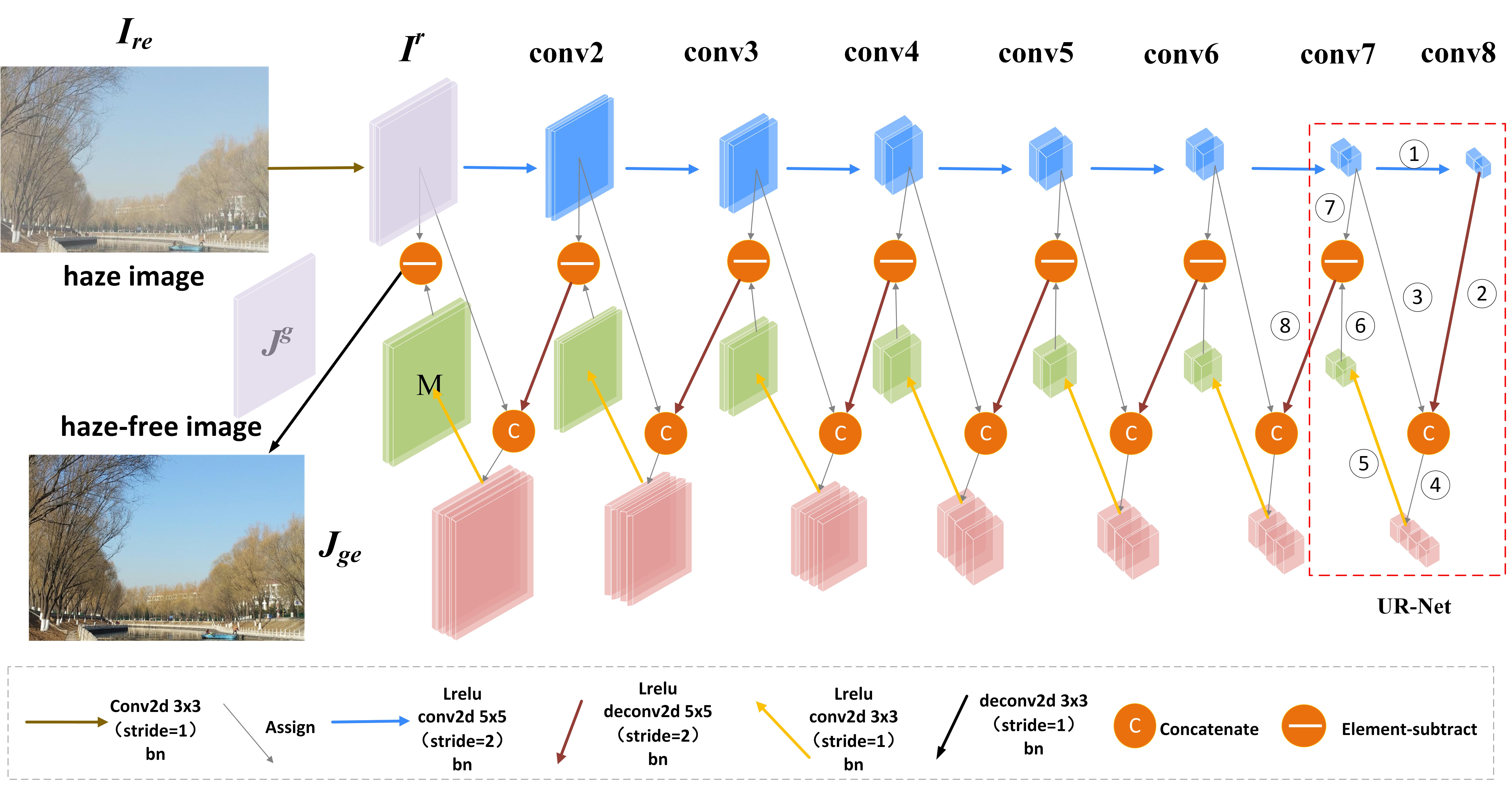}
	\caption{The designed generator (UR-Net-7) of the proposed input-size flexibility conditional generative adversarial network.}
	\label{fig:generator}
\end{figure*} 

Through log transformation, Eq.(\ref{eq:2}) can be transformed to the following form:
\begin{equation}
\log(J_{re}(x)-\alpha) = \log(I_{re}(x) - \alpha)-\log(t(x)),
\label{eq:6}
\end{equation}
By setting $J^g(x)=\log(J_{re}(x)-\alpha)$, $I^r(x)=\log(I_{re}(x)-\alpha)$ and $M(x)=\log(t(x))$, Eq.(\ref{eq:6}) can be rewritten as follows:
\begin{equation}
J^g(x) = I^r(x) - M(x),
\label{eq:5}
\end{equation}

In this paper, Encoder-decoder idea is used to realize dehazing task. According to Eq.(\ref{eq:5}), if we assume that $I^r$ is one layer output of the encoder network with input $I_{re}$, and $J^g$ is one layer output of the decoder network. We can obtain $J_{ge}$ according to the following rule: $I^r=\log(I_{re}-\alpha)\Rightarrow \alpha=I_{re}-\exp(I^r)$ then $J^g=\log(J_{ge}-\alpha)\Rightarrow J_{ge}=I_{re}-[exp(I^r)-exp(J^g)]$. In the following, we design our framework based on this observation and Eq.(\ref{eq:5}).

\subsection{The Generator of the Proposed Input-size Flexibility cGAN}
\label{se:3-2}

As derived in Eq.(\ref{eq:5}), residual idea is an important component of our generator. An U-connection residual network (UR-Net) is designed for single image dehazing, and the whole generator is the iteration of UR-Net between two adjacent convolutional layers. Fig. \ref{fig:generator} is the framework of our designed generator.

The unit of the red dotted rectangle in Fig. \ref{fig:generator} is an example of the designed UR-Net. Step \circled{1} is a convolutional layer with kernel of 5 $\times$ 5 and stride 2. Suppose that the shape of conv7 is $(1,c_7,h_7,w_7)$ and the shape of conv8 is $(1,c_8,h_8,w_8)$, then we have $h_8=\lceil \frac{h_7}{2}\rceil$, $w_8=\lceil \frac{w_7}{2}\rceil$, where $\lceil \cdot \rceil$ is an up-round symbol. This is implemented by using the "same" padding operation in TensorFlow.  Step \circled{2} is a de-convolutional layer with a kernel of 5 $\times$ 5 and stride 2. Suppose that the output shape of step \circled{2} is $(1,c_8^o,h_8^o,w_8^o)$, then we have $h_8^o=h_7$, $w_8^o=w_7$ by setting $h_8^o=\frac{h_8+1}{2}$, $w_8^o=\frac{w_8+1}{2}$. Next, we concatenate the output of step \circled{2} and conv7 in the channel dimension, the output of step \circled{4} is the concatenated result. This is the idea of U-Net for the purpose of fine information recovery. In order to realize residual learning between the input (conv7 in this example) and output of the penultimate layer of UR-Net, we need to ensure that the output channel dimension of the penultimate layer equals to the input. Thus, in step \circled{5}, a convolutional layer with kernel of 3 $\times$ 3 and stride 1 is used to reduce the channel dimension of step \circled{4}, and the output size of step \circled{5} is equal to its input size by using the "same" padding operation in TensorFlow. Finally, the residual can be obtained by the subtraction between conv7 and the output of step \circled{5}. Moreover, batch normalization (bn) \cite{34IoffeS15} is used to each convolutional and de-convolutional layer in our framework for the purpose of fast convergence. The activation function of the last layer is tanh($\cdot$), other layers are leak ReLU (Lrelu) and the value of leak is set to 0.2.

In order to provide noise to realize conditional input, dropout operation is used at both training and test stage after the de-convolutional layers corresponding to conv6, conv7 and conv8. The dropout rate is set as 0.5. In Fig. \ref{fig:generator}, the height $h_i$ and width $w_i$ of each conv{$i$} is related to the height $h$ and width $w$ of an input image, the calculation formulas are $h_i = \frac{h+2^i-1}{2^i}$ and $w_i = \frac{w+2^i-1}{2^i}$. The designed generator is an encoder-decoder structure, conv1 to conv8 form the encoder, the other parts form the decoder.

For convenience, in the following, UR-Net-$K$ is used to indicate that the number of UR-Net structure in the generator is $K$ (e.g., the generator of Fig. \ref{fig:generator} has 7 UR-Net structure, thus we call it UR-Net-7). At the same time, UR-Net-$K^*$ is used to represent that there is no subtraction process in the last UR-Net of the generator (the last UR-Net is located at the last de-convolutional layer).

The purpose of generator ($G$) is generating a haze-free output image $J_{ge}$ based on the input haze image $I_{re}$ and random noise $Z$, i.e., $G: \{I_{re},Z\} \rightarrow J_{ge}$.

\subsection{The Discriminator of the Proposed Input-size Flexibility cGAN}
\label{se:3-3}

\begin{figure*}
	\begin{center}
		\includegraphics[width=0.9\linewidth]{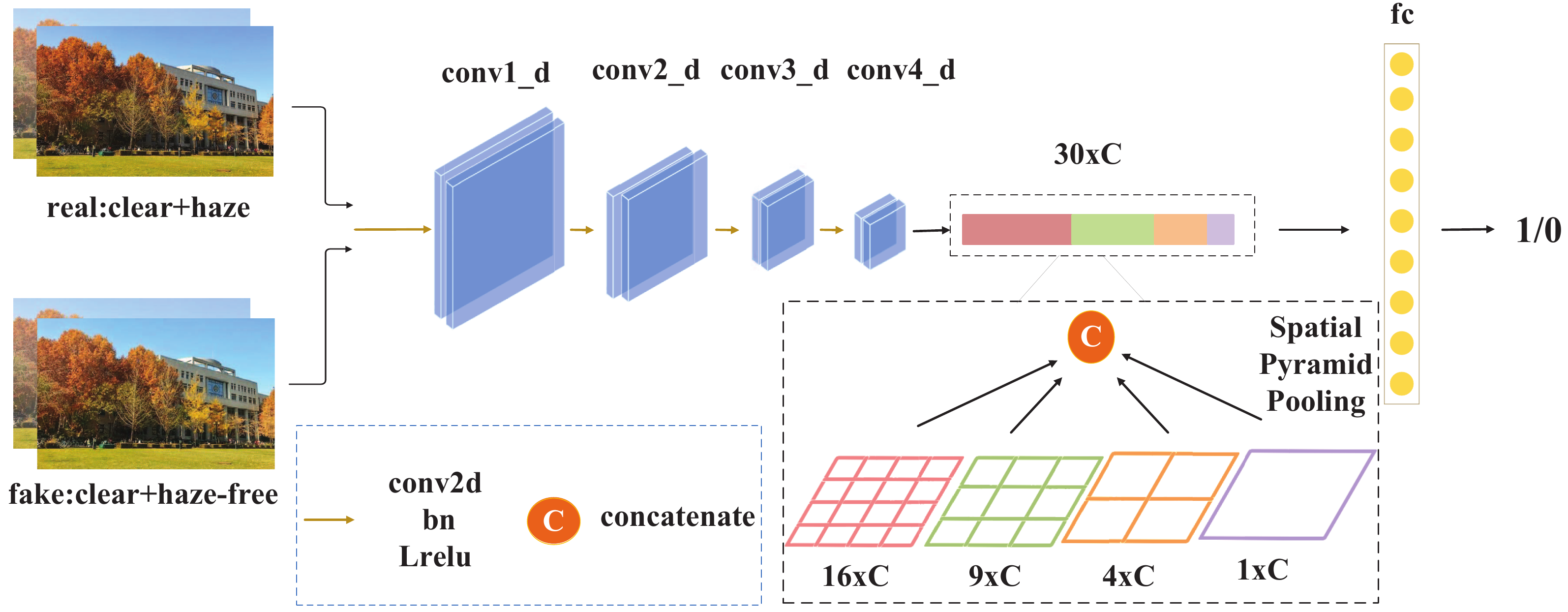}
	\end{center}
	\caption{The designed discriminator of the proposed input-size flexibility conditional generative adversarial network.}
	\label{fig:discriminator}
\end{figure*}

The discriminator $D$ is an important part of our proposed input-size flexibility cGAN model, which is used to discriminate the input sample is a "real image" ($J_{re}$) or a "generated image" ($J_{ge}$). As shown in Fig. \ref{fig:discriminator}, it consists of an input layer, 4 convolutional layers, a spatial pyramid pooling (SPP) \cite{35HeZR015} layer and a fully convolutional layer (fc). One image of the input layer is the concatenation of real clear image and real haze image across the channel dimension, another one is the concatenation of real clear image and the generated haze-free image across the channel dimension. The first three convolutional layers have a convolutional operation with kernel of 5 $\times$ 5 and stride 2, the last convolutional layer have a convolutional operation with kernel of 5 $\times$ 5 and stride 1. The SPP layer is designed to pool different sizes of input feature maps into vectors of the same length (the level of SPP is set as 4), thus training with input-size flexibility can be realized. The fc layer is a classifier to discriminate whether the input sample is real or fake (generated).

\subsection{Multi-loss Function}
\label{se:3-4}

The idea of multi-loss function optimization is widely used in various CNN-based systems, which is proved effective in different kinds of applications. It is also used in our framework. Next, we will define them one-by-one.

To ensure that $I^r=\log(I_{re}-\alpha)$ and $J^g=\log(J_{ge}-\alpha)$, we need to constrain $I_{re}-\exp(I^r)=J_{ge}-\exp(J^g)$. Thus, we define a consistency loss as follows:
\begin{equation}
\mathcal{L}_{Consistency}(G) = ||I_{re}-\exp(I^r)-J_{ge}+\exp(J^g)||_1
\label{eq:consistency}
\end{equation}
This consistency loss function is to ensure that the transformations of $I_{re}$ and $J_{ge}$ in the network are approximated to the log transformation with parameter $\alpha$, which is novel and important for our framework. Instead of learning the parameter of $\alpha$, by this consistency loss, a convolutional layer is developed to estimate the transformation of $\log(I_{re}-\alpha)$ and the inverse transformation of $\log(J_{ge}-\alpha)$, respectively. 

Then, we adopt the general cGAN loss function \cite{31IsolaZZE17} in our model, which is defined as follows:
\begin{equation}
\begin{split}
\mathcal{L}_{cGAN}(G,D)=&\mathbb{E}_{I_{re},J_{re}}[\log D(I_{re},J_{re})]+\\&\mathbb{E}_{I_{re},Z}[\log (1-D(I_{re},G(I_{re},Z)))]
\end{split}
\label{eq:cgan}
\end{equation}

At the training stage, the generator $G$ is trained to produce outputs that cannot be
distinguished as "fakes" by the discriminator $D$, and $D$ is trained to distinguish the generated example as "fakes". Thus, $G$ tries to minimize objective (\ref{eq:cgan}) against an adversarial $D$ that tries to maximize it, i.e., $G^*=arg \min_G\max_D \mathcal{L}_{cGAN}(G;D)$. In the last term of (\ref{eq:cgan}), minimizing $G$ is equivalent to maximizing $\log(D(I_{re},G(I_{re},Z)))$, which is adopted at the implementation stage.

Because the $L_1$ loss function can constrain the output of the generator absoutely equal to the expected output thus reduce the blur. We also introduce it as one of our loss functions, as follows:
\begin{equation}
\mathcal{L}_{L_1}(G)=\mathbb{E}_{I_{re},J_{re},Z}[||J_{re}-G(I_{re},Z)||_1]
\end{equation}

Moreover, perception-driving losses are verified effective in various image restoration tasks. Thus, in order to make the generated haze-free images have a good visual effect, we adopt SSIM and PSNR loss to construct our perception losses. In our model, the calculation formula of SSIM is the same as \cite{36ZhaoGFK17}. PSNR is defined as:
\begin{equation}
PSNR(J_{re},J_{ge})=10\cdot log_{10}(\frac{(\max(J_{re})-\min(J_{re}))^2}{MSE(J_{re},J_{ge})}),
\end{equation}
where $MSE(J_{re},J_{ge})$ is the mean of $(J_{re}-J_{ge})^2$ and can be formulated as $MSE(J_{re},J_{ge})=mean((J_{re}-J_{ge})^2)$.

According to the above formula, SSIM and PSNR losses are defined as follows:
\begin{equation}
\mathcal{L}_{SSIM}(G) = 1-SSIM(J_{re},J_{ge}),
\end{equation}
\begin{equation}
\mathcal{L}_{PSNR}(G) = 1-\frac{PSRN(J_{re},J_{ge})}{thresh},
\end{equation}
where $thresh$ is a threshold that is set to 40 in our experiments.

Finally, the overall loss function of our model is defined as follows:
\begin{equation}
\begin{split}
\mathcal{L} =& \mathcal{L}_{Consistency}(G) + \lambda_1\mathcal{L}_{cGAN}(G,D)+\lambda_2\mathcal{L}_{L_1}(G)+\\&\lambda_3\mathcal{L}_{SSIM}(G)+\lambda_4\mathcal{L}_{PSNR}(G)+\lambda||w||_2^2,
\end{split}
\label{eq:loss}
\end{equation}
where $\lambda_1$, $\lambda_2$, $\lambda_3$ and $\lambda_4$ are weights of their corresponding loss functions, which are set to 1, 100, 100 and 100 in our experiments, respectively. The final goal is to minimize (\ref{eq:loss}). The last term is only used to multi-scale training stage (Section \ref{se:3-5}), in which $w$ is the weights of the generator and $\lambda$ is the weight of this term. 

\subsection{Multi-scale Generator Fusion}
\label{se:3-5}
\begin{figure*}
	\begin{center}
		\includegraphics[width=0.9\linewidth]{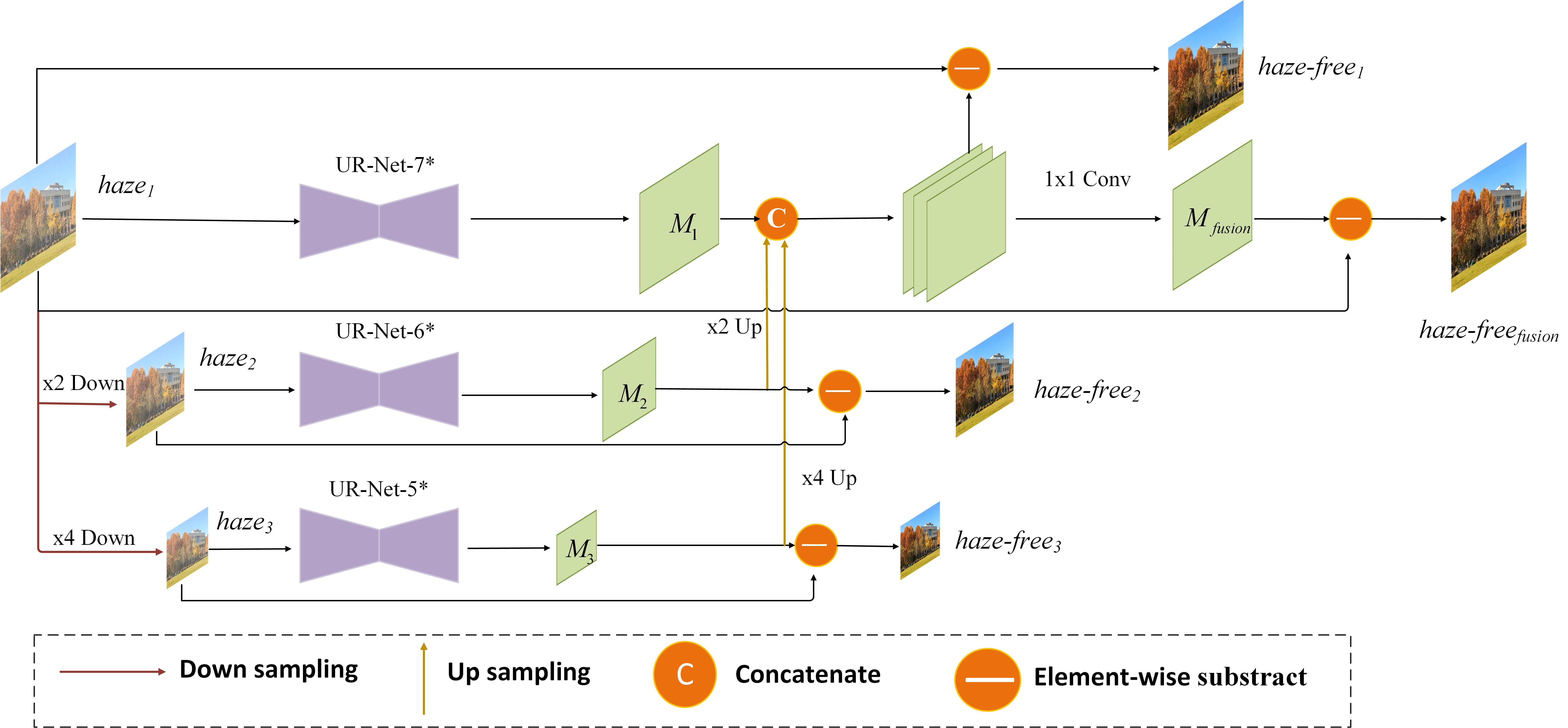}
	\end{center}
	\caption{The fusion model of multi-scale generator.}
	\label{fig:fusion}
\end{figure*} 

Multi-scale fusion is verified effective in image dehazing by Zhang and Tao \cite{6abs-1906-04334}. In their model, a Gaussian pyramid architecture with a late fusion module is designed to fuse different estimated feature maps.  Here, the Gaussian pyramid architecture is also used to realize multi-scale generator fusion model, as shown in Fig. \ref{fig:fusion}. This fusion model aims to show the generalization of our model to multi-scale framework. It should be noted that FAMED-Net\cite{6abs-1906-04334} is trained without adversarial, our multi-scale generator is trained based on our cGAN framework. The input of the generator includes one haze image (haze$_1$) and the corresponding down-sampled images ($\frac{1}{2}$ scale (haze$_2$) and $\frac{1}{4}$ scale (haze$_3$)). The output of the generator includes 4 haze-free images (haze-free$_1$, haze-free$_2$, haze-free$_3$, and haze-free$_{fusion}$ in Fig. \ref{fig:fusion}), which corresponds to the original scale of input haze image, $\frac{1}{2}$ scale, $\frac{1}{4}$ scale, and multi-scale fused output. The module of multi-scale fusion is performed based on the concatenation of haze maps ($M_1$, $M_2$, $M_3$ in Fig. \ref{fig:fusion}) of original scale, 2$\times$ up-sampling of $\frac{1}{2}$ scale, 4$\times$ up-sampling of $\frac{1}{4}$ scale. The fused haze map ($M_{fusion}$) is obtained after applying a convolutional layer with 1$\times$1 kernel to the concatenated haze maps. In this fusion model, the down-sampling and up-sampling operations are performed with bicubic interpolation. The generators of original scale, $\frac{1}{2}$ down-sampling scale and $\frac{1}{4}$ down-sampling scale of haze images are UR-Net-7$^*$, UR-Net-6$^*$ and UR-Net-5$^*$, respectively.

The discriminator is also vital for the fusion generator. Because the designed discriminator is input-size flexibility, thus we have two alternative of discriminator for the fusion generator, i.e., with and without sharing parameters for each output of the generator. Although sharing parameters of discriminator can reduce the model size, it can not reduce the number of computations. In order to enhance the discriminant ability of this fusion generator, we directly adopt the discriminator model without sharing parameters, i.e., each output of the fusion generator is discriminated by different discriminators.

For the loss function of the fusion generator, we apply objective (\ref{eq:loss}) to each output of the generator and corresponding discriminator.

\subsection{Model Training}
\label{se:3-6}

The model is implemented based on TensorFlow, and is trained with minibatch SGD (Stochastic Gradient Descent). The Adam solver \cite{37KingmaB14} with a learning rate of 0.0002 and momentum parameters $\beta_1=0.5$, $\beta_2=0.999$ is applied to optimize our model. All parameters are trained from scratch and the batch size is set as 1. The hyper-parameter $\lambda$ in objective (\ref{eq:loss}) is set as 0.001. In order to better maintain the convergence balance between the generator and the discriminator, we update parameters of the discriminator once every 4 iterations. Because our model is input-size flexibility, we can train this model by images with different sizes to obtain a better haze-free image. However, training with different sizes of input is slow, thus the model is first trained with fixed size of input images and then fine-tuned based on different sizes of images.

\section{Experiments}
\label{se:4}

We conduct experiments on the dataset of intelligent traffic video image enhancement processing competition of ICIG 2019 (we call it ICIG2019 for convenience) and the large scale REalistic Single Image DEhazing (RESIDE) dataset \cite{4LiRFTFZW19} for single image dehazing.
 
The \textbf{ICIG2019} dataset contains 5500 clear images of real scene and corresponding synthetic haze ones. The training and validation sets contain 5000 and 500 image pairs, respectively. In the experiments, we use the training set to train our models and use the validation set to test the trained models. Ablation studies are conducted on this dataset. 

The \textbf{RESIDE} dataset is one of the largest single image dehazing datasets, which contains 110,500 synthetic hazy indoor images (ITS) and 313,950 synthetic hazy outdoor images (OTS) in the training set. The synthetic objective testing set (SOTS) contains 500 indoor images and 500 outdoor images. The hybrid subjective testing set (HSTS) contains 10 real-world images and 10 synthetic images. In the training dataset, each clear image corresponds to multiple haze images of different concentrations. For each clear image, we randomly select a corresponding haze image from the training samples to form our training set.

We use 4 evaluation metrics that have been realized in the skimage package of python to evaluate the performance of single image dehazing, which are MSE (the smaller the better $\downarrow$), normalization root mean-squared error (NRMSE) (the smaller the better $\downarrow$), PSNR (the larger the better $\uparrow$) and SSIM (the larger the better $\uparrow$). We also re-test the compared methods by running the corresponding released models. All the results reported for the compared methods and our methods are evaluated using the standard evaluation interface of python for a fair comparison.

\subsection{Ablation Studies}
\label{se:4-1}

The experiments of ablation study is conducted on the ICIG2019 dataset. We first verify the effectiveness of each loss function in Eq.(\ref{eq:loss}) based on the model of UR-Net-7 (Fig. \ref{fig:generator}) and the input image with size of 256$\times$256. The maximum training epoch is set as 16. Based on the $\mathcal{L}_{Base}=\mathcal{L}_{Consistency}(G)+\mathcal{L}_{cGAN}(G,D)$ loss function in our framework, we verify the combinations of $\mathcal{L}_{Base}+\mathcal{L}_{L_1}$, $\mathcal{L}_{Base}+\mathcal{L}_{SSIM}$, $\mathcal{L}_{Base}+\mathcal{L}_{PSNR}$, $\mathcal{L}_{Base}+\mathcal{L}_{L_1}+\mathcal{L}_{SSIM}$ (without $\mathcal{L}_{PSNR}$), $\mathcal{L}_{Base}+\mathcal{L}_{L_1}+\mathcal{L}_{PSNR}$ (without $\mathcal{L}_{SSIM}$), $\mathcal{L}_{Base}+\mathcal{L}_{SSIM}+\mathcal{L}_{PSNR}$ (without $\mathcal{L}_{L_1}$), and $\mathcal{L}_{Base}+\mathcal{L}_{L_1}+\mathcal{L}_{SSIM}+\mathcal{L}_{PSNR}$ ($\mathcal{L}$). The experimental results are shown in Table \ref{tab:1}.

\begin{table}
	\caption{Results of Different Losses on The ICIG2019 Dataset.}
	\setlength{\tabcolsep}{1.5mm}{
	\begin{center}
		\begin{tabularx}{8.0cm}{lcccc}
			\hline
			loss & MSE $\downarrow$ & NRMSE $\downarrow$ & PSNR $\uparrow$ & SSIM $\uparrow$ \\
			\hline\hline
			$\mathcal{L}_{Base}$ & 637.6 & 0.168 & 21.54 & 0.789 \\
			$\mathcal{L}_{Base}+\mathcal{L}_{L_1}$ & 382.8 & 0.135 & 23.21 & 0.866 \\
			$\mathcal{L}_{Base}+\mathcal{L}_{SSIM}$ & 383.3 & 0.135 & 23.26 & 0.894 \\
			$\mathcal{L}_{Base}+\mathcal{L}_{PSNR}$ & 336.3 & 0.126 & 24.03 & 0.890 \\
			without $\mathcal{L}_{PSNR}$ & 321.1 & 0.123 & 24.15 & 0.899 \\
			without $\mathcal{L}_{SSIM}$ & 299.2 & 0.119 & 24.45 & 0.898 \\
			without $\mathcal{L}_{L_1}$ & 337.8 & 0.126 & 23.87 & 0.902 \\
			$\mathcal{L}$ & \textbf{287.9} & \textbf{0.116} & \textbf{24.58} & \textbf{0.904} \\
			\hline
		\end{tabularx}
	\end{center}
	}
	\label{tab:1}
\end{table}

From Table \ref{tab:1}, it can be seen that the best performance is obtained when all the losses are used. The performances of $\mathcal{L}_{Base}+\mathcal{L}_{L_1}$, $\mathcal{L}_{Base}+\mathcal{L}_{SSIM}$ and $\mathcal{L}_{Base}+\mathcal{L}_{PSNR}$ are much better than the performance of $\mathcal{L}_{Base}$, which verified the effectiveness of each loss function after combined with $\mathcal{L}_{Base}$. The performance without $\mathcal{L}_{PSNR}$ is much better than the performances of $\mathcal{L}_{Base}+\mathcal{L}_{L_1}$ and $\mathcal{L}_{Base}+\mathcal{L}_{SSIM}$, the performance without $\mathcal{L}_{SSIM}$ is much better than the performances of $\mathcal{L}_{Base}+\mathcal{L}_{L_1}$ and $\mathcal{L}_{Base}+\mathcal{L}_{PSNR}$, and the performance without $\mathcal{L}_{L_1}$ is much better than the performances of $\mathcal{L}_{Base}+\mathcal{L}_{SSIM}$ and $\mathcal{L}_{Base}+\mathcal{L}_{PSNR}$, which verified the effectiveness of combining any two loss functions with $\mathcal{L}_{Base}$. Moreover, we notice that the values of MSE and PSNR of $L_{Base} + L_{PSNR}$ are much better than $L_{Base} + L_{L_1}$ and $L_{Base} + L_{SSIM}$, which shows that the proposed PSNR loss is much better than the $L_1$ loss and $L_{SSIM}$ loss when they combined with $L_{Base}$, respectively.

The second ablation experiments are the generator network with fixed sizes of input images and fine-tuned with input-size flexibility images, which aims to verify the effectiveness of input-size flexibility. The experimental results are shown in Table \ref{tab:2}.

\begin{table}
	\caption{Results of Different Input Sizes With and Without Input-size Flexibility Fine-tuning on The ICIG2019 Dataset.}
	\setlength{\tabcolsep}{1.5mm}{
	\begin{center}
		\begin{tabularx}{8.0cm}{lcccc}
			\hline
			Training mode & MSE $\downarrow$ & NRMSE $\downarrow$ & PSNR $\uparrow$ & SSIM $\uparrow$ \\
			\hline\hline
			256$\times$256 & 287.9 & 0.116 & 24.58 & 0.904 \\
			256$\times$256 + IFF & 245.1 & 0.108 & 25.04 & 0.905 \\
			368$\times$544 & 300.8 & 0.118 & 24.53 & 0.898 \\
			368$\times$544 + IFF & \textbf{219.6} & \textbf{0.101} & 25.58 & 0.905 \\
			512$\times$512 & 317.1 & 0.116 & 24.67 & 0.891 \\
			512$\times$512 + IFF & 223.4 & \textbf{0.101} & \textbf{25.67} & \textbf{0.906} \\
			\hline
		\end{tabularx}
	\end{center}
	}
	\label{tab:2}
\end{table}

In Table \ref{tab:2}, the IFF is the abbreviation of input-size flexibility fine-tuning. The training mode indicates the sizes of training input. The test results are based on the mode of input-size flexibility, i.e., the output size of an image equals to the size of the input image. From Table \ref{tab:2}, we can see that with the input-size flexibility fine-tuning, better performances can be obtained. Moreover, the best MSE is obtained with the training mode of 368$\times$544 + IFF, the best PSNR and SSIM are obtained with the training mode of 512$\times$512 + IFF. The size of 368$\times$544 is the mean size of the training images (368 is the mean of heights and 544 is the mean of widths). Moreover, we can see that when IFF is not used, the best MSE is obtained by the model trained with input size of 256$\times$256. These experimental results show that if IFF is used, the process of pre-training with larger input size can lead to higher PSNR and SSIM, otherwise, the model trained with smaller input size can lead to a smaller MSE. However, both pre-training and fine-tuning are complex processes. The conclusion that can be determined from the experiments is that the performances with IFF are better than the performances without IFF. Other conclusions may be related to the experimental parameter settings.

\subsection{Comparison With the State-of-the-Art Methods}
\label{se:4-2}

We compare the proposed method with the state-of-the-art CNN-based dehazing methods. They are AOD-Net \cite{6LiPWXF17}, MSCNN \cite{7RenLZPC016}, GMAN \cite{9LiuXAWC19}, DCPDN \cite{100004P18a}, De-cGAN \cite{12LiPLT18}, GFN \cite{13Ren0ZPC0018}, and recently proposed FAMED-Net \cite{6abs-1906-04334}. The comparison results of ICIG2019 dataset are shown in Table \ref{tab:3}.

\begin{table}
	\caption{Comparison With The State-of-the-art Methods on The Validation of ICIG2019 Dataset.}
	 \setlength{\tabcolsep}{1.5mm}{
	\begin{center}
		\begin{tabularx}{8.4cm}{lcccc}
			\hline
			Methods & MSE $\downarrow$ & NRMSE $\downarrow$ & PSNR $\uparrow$ & SSIM $\uparrow$ \\
			\hline\hline
			MSCNN \cite{7RenLZPC016} & 1292 & 0.250 & 17.33 & 0.810 \\
			DCPDN \cite{100004P18a} & 971.2 & 0.218 & 19.06 & 0.848 \\
			GFN \cite{13Ren0ZPC0018} & 766.7 & 0.176 & 20.96 & 0.828 \\
			De-cGAN \cite{12LiPLT18} & 764.4 & 0.174 & 21.02 & 0.857 \\
			AOD-Net \cite{6LiPWXF17} & 646.8 & 0.175 & 20.73 & 0.868 \\
			GMAN \cite{9LiuXAWC19} & 290.2 & 0.118 & 24.37 & 0.887 \\
			GMAN fine-tuned & 287.1 & 0.118 & 24.43 & 0.891 \\
			FAMED-Net \cite{6abs-1906-04334} & 249.5 & 0.107 & 25.17 & 0.909 \\
			\hline
			UR-Net-7  & \textbf{223.4} & \textbf{0.101} & \textbf{25.67} & 0.906\\
			Multi-scale cGAN & \textbf{213.6} & \textbf{0.099} & \textbf{25.89} & \textbf{0.912}\\
			\hline
		\end{tabularx}
	\end{center}
	}
	\label{tab:3}
\end{table}


In Table \ref{tab:3}, the method of GMAN fine-tuned means the fine-tuned model of GMAN on the ICIG2019 dataset based on the pre-trained GMAN model. From Table \ref{tab:3}, we can see that the proposed UR-Net-7 is much better than the previous proposed methods for the evaluations of MSE, NRMSE and PSRN. The best SSIM is obtained by the proposed multi-scale cGAN, followed by the method of FAMED-Net. Moreover, we notice that after the GMAN model is fine-tuned (GMAN fine-tuned) on the ICIG2019 dataset, the performance are better than the GMAN without fine tuning (GMAN (SPL19)).

In these comparison methods, both AOD-Net and GMAN are input-size flexibility at the test stage. But they are not input-size flexibility at the training stage, one reason is that the batch-size of them is greater than 1 to obtain a good performance. The performance of them will drop a lot if the batch-size is set as 1 for input-size flexibility purpose at the training stage, because batch-normalization is adopted to realize good performance by using large batch-size. The FAMED-Net is designed based on the AOD-Net, it can be also changed to input-size flexibility mode at the test stage, because late fusion idea is adopted, better performance can be obtained. Different from these works, the proposed model is GAN-based input-size flexibility, which is input-size flexibility at both training and test stages. Moreover, the proposed multi-scale cGAN obtained the best single image dehazing performance based on the evaluations in this paper, which also proved the effectiveness of image late fusion. Different from previous fusion idea, the proposed fusion framework is based on cGAN, which is a cGAN fusion framework.

\begin{table}
	\caption{Comparison With The State-of-the-art Methods on The Outdoor of SOTS Dataset.}
	\setlength{\tabcolsep}{1.5mm}{
	\begin{center}
		\begin{tabularx}{8.4cm}{lcccc}
			\hline
			Methods & MSE $\downarrow$ & NRMSE $\downarrow$ & PSNR $\uparrow$ & SSIM $\uparrow$ \\
			\hline\hline
			MSCNN \cite{7RenLZPC016} & 812.2 & 0.202 &20.02 & 0.880 \\
			DCPDN \cite{100004P18a} & 828.1  & 0.204 & 19.93 & 0.858 \\
			GFN \cite{13Ren0ZPC0018} &  676.2& 0.172 & 21.47 & 0.849 \\
			De-cGAN \cite{12LiPLT18} & 611.1 & 0.160 & 21.96 & 0.868 \\
			AOD-Net \cite{6LiPWXF17} & 693.0 & 0.185  & 20.47 &0.899  \\
			FAMED-Net \cite{6abs-1906-04334} & 199.6  & 0.098 & 26.17 & \textbf{0.925} \\
			\hline
			UR-Net-7  & \textbf{160.5} & \textbf{0.089}  & \textbf{26.96} & 0.910\\
			Multi-scale cGAN & \textbf{152.3} & \textbf{0.086} & \textbf{27.28} & \textbf{0.925}\\
			\hline
		\end{tabularx}
	\end{center}
	}
	\label{tab:4}
\end{table}

\begin{table}
	\caption{Comparison With The State-of-the-art Methods on The Synthetic of HSTS Dataset.}
	\setlength{\tabcolsep}{1.5mm}{
	\begin{center}
		\begin{tabularx}{8.4cm}{lcccc}
			\hline
			Methods & MSE $\downarrow$ & NRMSE $\downarrow$ & PSNR $\uparrow$ & SSIM $\uparrow$ \\
			\hline\hline
			MSCNN \cite{7RenLZPC016} & 1164.2 & 0.233 & 18.47 & 0.813 \\
			DCPDN \cite{100004P18a} & 841.5 & 0.197 & 20.21 & 0.852 \\
			GFN \cite{13Ren0ZPC0018} & 527.6 & 0.147 & 22.83 & 0.887 \\
			De-cGAN \cite{12LiPLT18} & 498.5 & 0.145 & 22.85 & 0.869 \\
			AOD-Net \cite{6LiPWXF17} & 711.1 & 0.181 & 20.56 & 0.887 \\
			FAMED-Net \cite{6abs-1906-04334} & 168.9 & 0.089 & 26.68 & \textbf{0.922} \\
			\hline
			UR-Net-7  & \textbf{146.3} & \textbf{0.083} & \textbf{27.04} & 0.908\\
			Multi-scale cGAN & \textbf{105.9} & \textbf{0.071} & \textbf{28.38} & 0.919\\
			\hline
		\end{tabularx}
	\end{center}
	}
	\label{tab:5}
\end{table}

\begin{figure*}
	\begin{center}
		\includegraphics[width=1.0\linewidth]{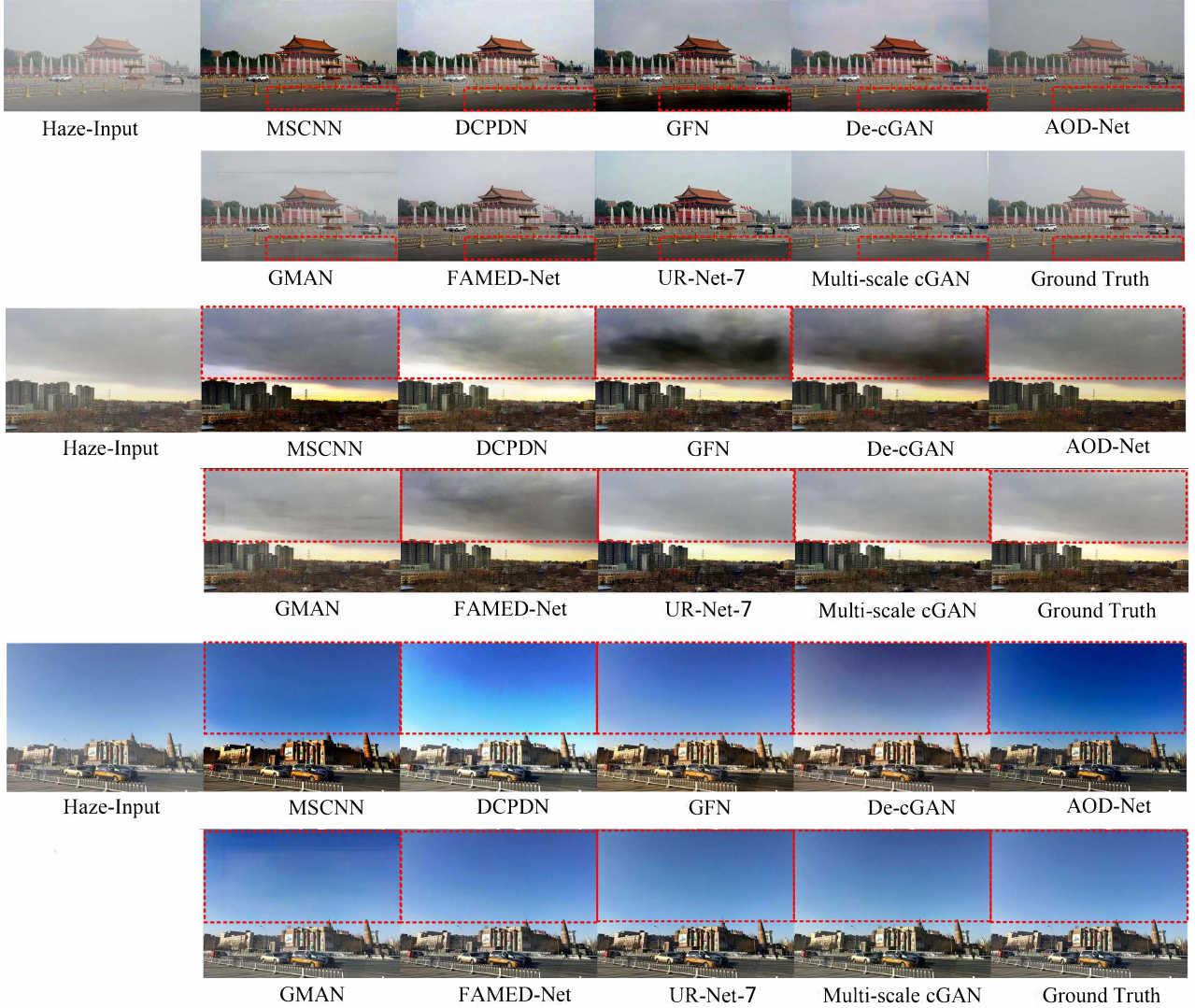}
	\end{center}
	\caption{Subjective comparisons between the proposed methods and the most related state-of-the-art methods on synthetic hazy images from ICIG2019 validation set. Best viewed in color.}
	\label{fig:show_icig}
\end{figure*}

\begin{figure*}
	\begin{center}
		\includegraphics[width=1.0\linewidth]{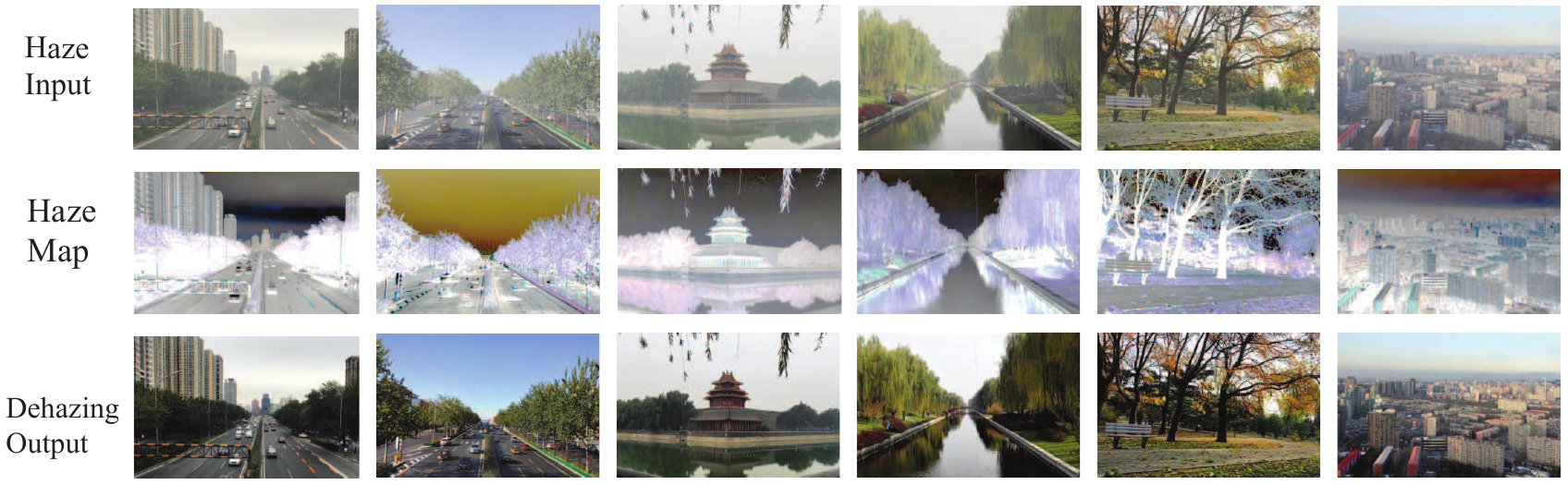}
	\end{center}
	\caption{The visualizations of haze map $M$ and corresponding dehazed results of UR-Net-7.}
	\label{fig:hazemap}
\end{figure*}

\begin{figure*}
	\begin{center}
		\includegraphics[width=1.0\linewidth]{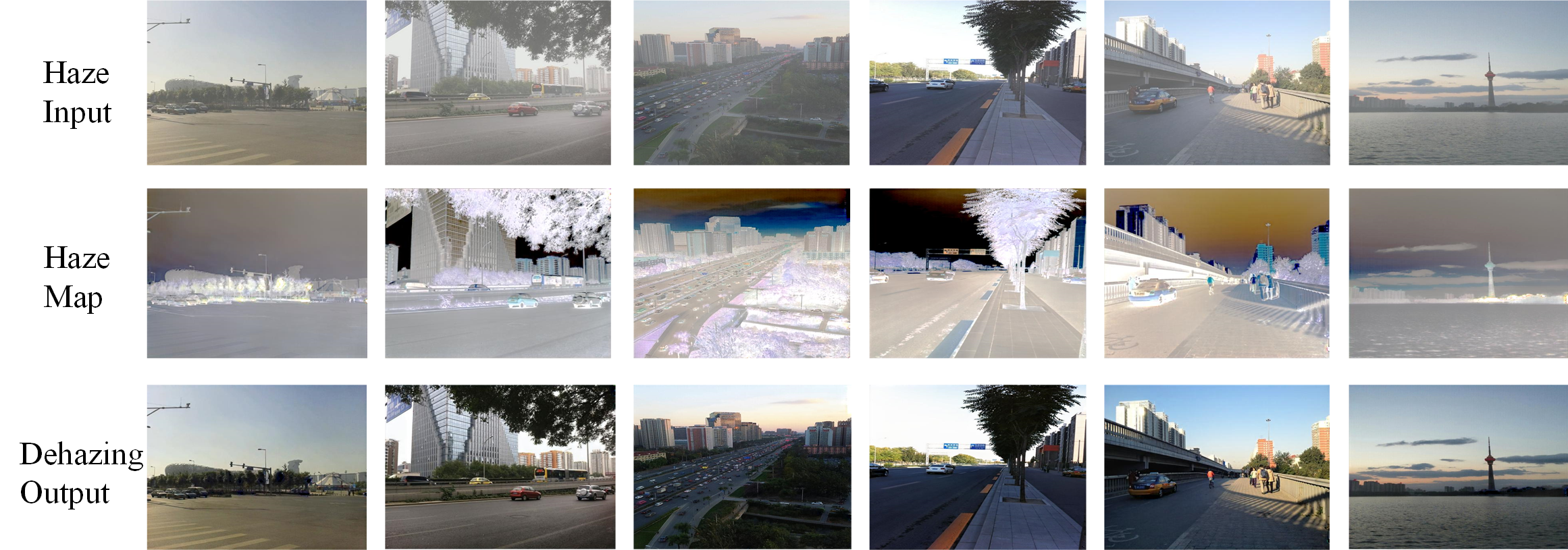}
	\end{center}
	\caption{The visualizations of haze map $M_{fusion}$ and corresponding dehazed results of Multi-scale cGAN.}
	\label{fig:hazemap-multi}
\end{figure*}

Table \ref{tab:4} and Table \ref{tab:5} are the comparisons of the outdoor of SOTS and HSTS on the RESIDE dataset. It can be seen that the proposed multi-scale cGAN obtains the best results for the evaluations of MSE, NRMSE and PSNR. For the evaluation of SSIM, the best value is obtained by the method of FAMED-Net (both in Table \ref{tab:4} and Table \ref{tab:5}) and multi-scale cGAN (in Table \ref{tab:4}). Although the SSIM of the proposed multi-scale cGAN is less than the FAMED-Net 0.03\% in Table \ref{tab:5}, the MSE, NRMSE, and PSNR values of the proposed multi-scale cGAN are much higher than the FAMED-Net. In particular, the PSNR is 1.34dB higher.

Fig. \ref{fig:show_icig} is the subjective comparisons on synthetic hazy images from the ICIG2019 validation set. From these dehazed images, we can see that our methods (especially multi-scale cGAN) are relatively good for the ground, the clouds and the sky.

Table \ref{tab:6} is the comparisons of the indoor of SOTS on the RESIDE dataset. According to Table \ref{tab:6}, we can see that the best SSIM is obtained by the FAMED-Net. However, the best values of MSE, NRMSE, and PSNR are obtained by the proposed UR-Net-7 and multi-scale cGAN.

\begin{table}
	\caption{Comparison With The State-of-the-art Methods on The Indoor of SOTS Dataset.}
	\setlength{\tabcolsep}{1.5mm}{
	\begin{center}
		\begin{tabularx}{8.4cm}{lcccc}
			\hline
			Methods & MSE $\downarrow$ & NRMSE $\downarrow$ & PSNR $\uparrow$ & SSIM $\uparrow$ \\
			\hline\hline
			MSCNN \cite{7RenLZPC016} & 2097.5 & 0.383 & 16.00 & 0.780  \\
			AOD-Net \cite{6LiPWXF17} & 1144.8 & 0.271  & 19.07 & 0.824 \\
			GFN \cite{13Ren0ZPC0018} & 443.0&0.175 &22.48 & 0.888 \\
			FAMED-Net \cite{6abs-1906-04334} & 361.4  & 0.153 & 23.63 & \textbf{0.901} \\
			\hline
			UR-Net-7 & \textbf{274.1} & \textbf{0.139} & \textbf{24.42} & 0.881 \\
			Multi-scale cGAN & \textbf{265.3} & \textbf{0.132} & \textbf{24.56} & 0.900 \\
			\hline
		\end{tabularx}
	\end{center}
	}
	\label{tab:6}
\end{table}

\begin{table}
	\caption{Learnable Parameters and Time Spent of Different Methods.}
	\setlength{\tabcolsep}{1.5mm}{
	\begin{center}
		\begin{tabularx}{6.8cm}{lcc}
			\hline
			Methods & Params & Time(second) \\
			\hline\hline
			MSCNN \cite{7RenLZPC016} & 8,014& 0.04\\
			AOD-Net \cite{6LiPWXF17} & 1,833& 0.004\\
			De-cGAN \cite{12LiPLT18} & 1.23$\times 10^8$& 0.05\\
			DCPDN \cite{100004P18a} & 6.69$\times 10^7$& 0.04\\
			GFN \cite{13Ren0ZPC0018} & 514,415& 0.05\\
			FAMED-Net \cite{6abs-1906-04334} & 17,991 & 0.03\\
			\hline
			UR-Net-7 & 8.59$\times 10^7$ & 0.04\\
			Multi-scale cGAN & 2.06 $\times 10^8$& 0.1\\
			\hline
		\end{tabularx}
	\end{center}
	}
	\label{tab:time}
\end{table}
Table \ref{tab:time} summarizes the learnable parameters and time spent of different models based on the Tesla K40c GPU. The numbers of learnable parameters of our methods are larger than other methods. This is because our model includes discriminators. However, the discriminators are not used during test. The time spent of our UR-Net-7 is similar to most of other state-of-the-art learning-based methods. 

\subsection{Discussions}

\begin{figure*}
	\begin{center}
		\includegraphics[width=1.0\linewidth]{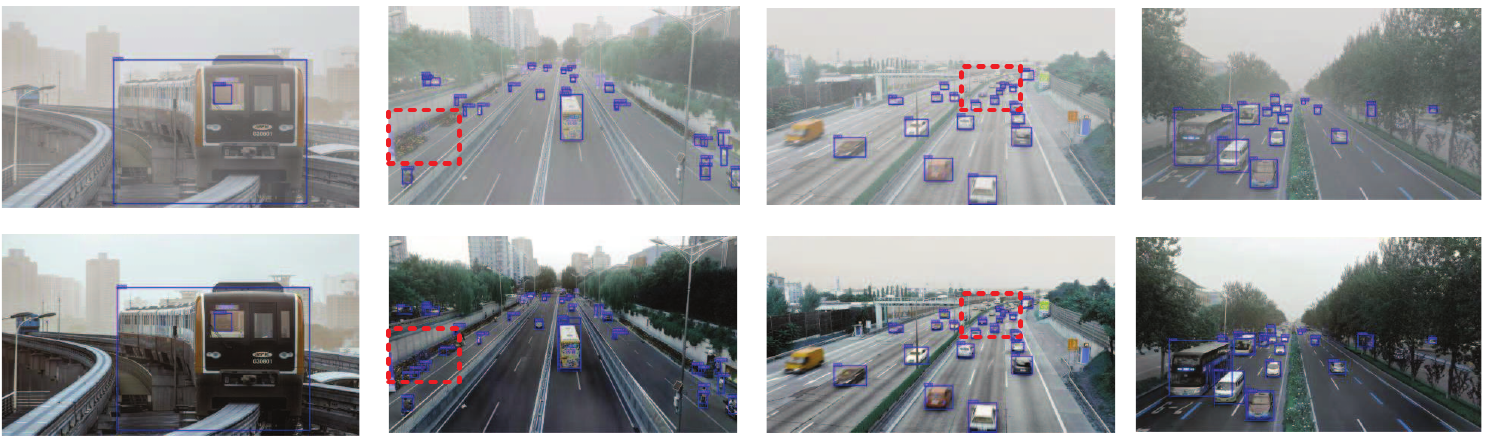}
	\end{center}
	\caption{The detection results before (the first line) and after (the second line) dehazing with UR-Net-7.}
	\label{fig:detection}
\end{figure*}

As analysis in Section \ref{se:3-1}, the general atmospheric scattering model can be simplified to Eq.(\ref{eq:5}). According to Eq.(\ref{eq:5}), one haze image ($I$) can be seen as a clear image ($J$) plus a content related noise image $M$, which is a general additive  noise model. For CNN-based image denoising or restoration, most of the noise models can be transformed into an additive noise model, e.g., the multiplicative noise model can be transformed into an additive noise model by logarithmic transformation. Thus, the proposed input-size flexibility cGAN is a general image restoration model.

The haze map ($M$) of Eq.(\ref{eq:5}) can be thought of as a kind of content related noise in a haze image. The visualizations of $M$ are shown in Fig. \ref{fig:hazemap}. The haze maps in Fig \ref{fig:hazemap} is the transformed results of $M$, which is same as the transformation of $J^g$, i.e., add 1 and multiply by the mean. From Fig. \ref{fig:hazemap}, it can be seen that the haze maps relate with the color, illumination and the concentration of haze, also the content of the corresponding haze images. Similar to the haze of real scenes, there is no specific rule for these generated haze maps. Moreover, the visualizations of $M_{fusion}$ for multi-scale cGAN are also shown in Fig. \ref{fig:hazemap-multi}, the characteristics of these haze maps are similar to those in Fig. \ref{fig:hazemap}.
 
Considering the applicability, image dehazing can usually be used to the preprocess step of other computer vision tasks. The proposed image dehazing algorithm can be used to assist object detection, as shown in Fig.\ref{fig:detection}, which is the comparison of object detection results before and after dehazing with the proposed UR-Net-7. The detection algorithm is SNIPER \cite{38SinghND18}, we only use the released code\footnote{https://github.com/mahyarnajibi/SNIPER} and the pre-trained model for detection. From the detection results of the two images in the middle of Fig.\ref{fig:detection}, we can see that more objects can be detected after dehazing with UR-Net-7 (the objects in the red rectangle).

\begin{table}
	\caption{Summary of Major Contributions for Performance Improvement Based on The Experimental Results of ICIG2019 Dataset.}
	\setlength{\tabcolsep}{1.5mm}{
	\begin{center}
		\begin{tabularx}{8.0cm}{lcccc}
			\hline
			Methods & MSE $\downarrow$ & NRMSE $\downarrow$ & PSNR $\uparrow$ & SSIM $\uparrow$ \\
			\hline\hline
			Baseline & 637.6 & 0.168 & 21.54 & 0.789  \\
			\quad+ Mulit-loss & 287.9 & 0.116 & 24.58 & 0.904\\
			\quad\quad+ IFF & 223.4 & 0.101 & 25.67 & 0.906\\
			\hline
		\end{tabularx}
	\end{center}
	}
	\label{tab:7}
\end{table}

\begin{figure*}
	\begin{center}
		\includegraphics[width=1.0\linewidth]{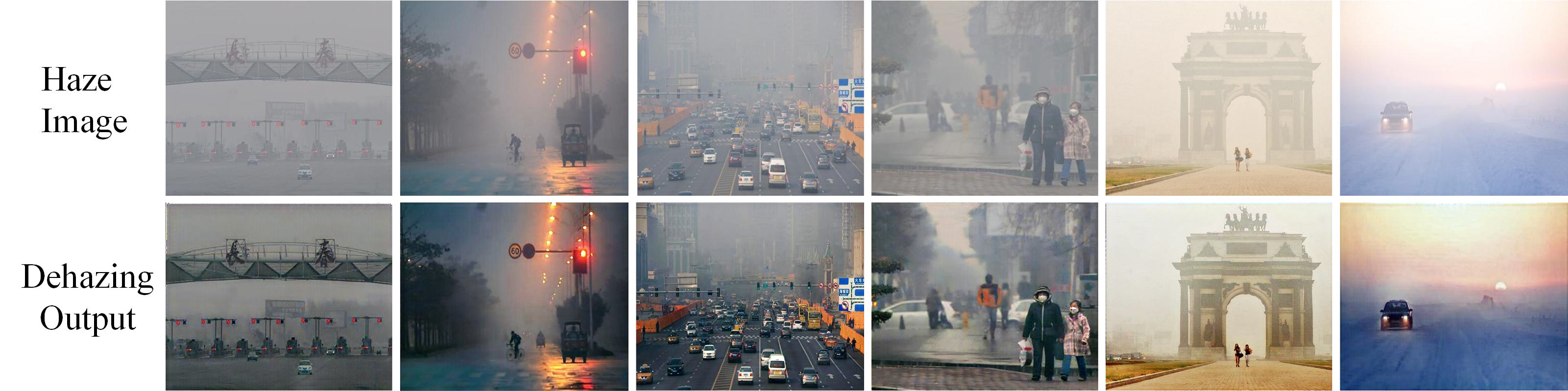}
	\end{center}
	\caption{The dehazing results for images of nighttime or low-light conditions based on the UR-Net-7 model trained on the RESIDE outdoor training set.}
	\label{fig:nightime}
\end{figure*}

Moreover, the UR-Net-7 model trained on the RESIDE outdoor training set is used to test haze images in nighttime or low-light conditions. Results are shown in Fig. \ref{fig:nightime}, it can be seen that the proposed model can be generalized to the haze scene under special conditions. 

Finally, the major contributions of our work for performance improvement based on the experimental results of ICIG2019 dataset are summarized in Table \ref{tab:7}. We can see that optimization with multi-loss functions greatly improves the performance of baseline, boosted 3.04 dB and 12.5\% for PSNR and SSIM evaluation metrics, respectively. Moreover, input-size flexibility fine-tuning (IFF) can further improve the PSNR about 1.09 dB. 
\section{Conclusions}
\label{se:5}

In this paper, an input-size flexibility cGAN with multi-loss function training model is developed for single image dehazing, experimental results proved the effectiveness of input-size flexibility and multi-loss function optimization. Moreover, a multi-scale image restoration fusion framework based on cGAN was proposed and verified for single image dehazing. Experimental results showed that we obtained the best single image dehazing performance on the ICIG2019 and RESIDE datasets. On the ICIG2019 dataset, the PSNR has been improved 0.5dB and 0.72dB for UR-Net-7 and Multi-scale cGAN compared with the FAMED-Net, respectively. On the outdoor of SOTS dataset, the PSNR has been improved 0.79dB and 1.11dB for UR-Net-7 and Multi-scale cGAN compared with the state-of-the-art methods, respectively.
Our basic idea is to realize image restoration based on Eq.(\ref{eq:5}), thus the proposed framework can be also used to other image restoration tasks, such as image denoising, image deblurring, and image fusion \cite{39GuoNCZMH19}. Future works could be focused on extending our methods to other image restoration tasks.

\printcredits

\bibliographystyle{cas-model2-names}

\bibliography{cas-refs}






\end{document}